"Some Comments on the Theory of Bose Condensation, with particular reference to the interiors of Stars"

J. Dunning-Davies,
Physics Department,
University of Hull,
Hull HU6 7RX,
England.

J.Dunning-Davies@hull.ac.uk

**Abstract.**

The possibility of the phenomenon of Bose condensation having a part to play in the discussion of neutron stars has been around for some time. Here the sorts of temperatures and densities that might be involved are discussed. Also, an alternative way of viewing the Bose condensation phenomenon is examined once more and, although found to lead to more accurate results in traditional examples, is found to have little numerical effect in astrophysical examples.

## 1. Introduction.

For some time now, the suggestion has been abroad that kaon condensation could have a role to play in the understanding of neutron stars [1]. In fact, the appearance of a Bose-Einstein condensate of charged pions in dense nuclear matter has been under discussion since the early seventies, as was noted in the above reference. However, as far as a possible Bose condensation occurring in a star or, indeed, the notion of a star being composed primarily of a Bose condensate, are concerned, questions of admissible condensation temperatures and corresponding number densities do not appear to have been addressed. This is easily rectified in the simple case of an ideal Bose gas by referring to an article on ideal relativistic Bose condensation dating back to 1965 [2] and leads to some interesting results.

However, it is noted also that there are problems with the traditional approach to the whole question of Bose-Einstein condensation. A totally different approach was investigated some years ago [3] but is revisited here to illustrate, once again, that at least some of the well-known difficulties of the usual approach to the problem may be avoided.

## 2. Theory of Bose Condensation.

In the usually accepted way, consider an ideal Bose gas with $\nu(E)dE$ single-particle quantum states in an energy range $dE$. Here $E$ excludes the rest energy $\varepsilon_o = mc^2$. If the number of particles in the lowest energy level is $N_1(\alpha,T)$ at temperature $T$, then the total number of particles in the system is

$$N - N_1(\alpha,T) = \int_0^\infty \nu(E) f(E,\alpha,T) dE, \quad (2.1)$$

where $\alpha = (\mu - \varepsilon_o)/kT$, $\mu$ being the chemical potential. As the temperature is lowered, the numerical value of $\alpha$, in the limit of infinite volume, decreases, and can in some cases become zero at a critical temperature $T_c > 0$. Hence, for a large volume, this condensation temperature may be defined by

$$\alpha < 0 \text{ for } T > T_c, \quad \alpha = 0 \text{ for } T \leq T_c. \quad (2.2)$$

If (2.2) is used in (2.1), a unique condition for $T = T_c$ is not obtained unless (2.2) is supplemented by

$$N_1(\alpha,T) << N \text{ for } T \geq T_c, \qquad (2.3)$$

where the equality sign is important. Below the condensation temperature, $N_1(\alpha,T)$ is a non-negligible fraction of $N$. Accordingly, $T_c$ follows from the relation

$$N = \int_0^\infty \nu(E) f(E,0,T_c) dE \qquad (2.4)$$

Here a previous argument has been followed [4] but a more rigorous foundation for these equations is possible [2].

For a relativistic gas, the density of states is

$$\nu(E) = (4\pi V/h^3 c^3)(E + \varepsilon_o)(E^2 + 2\varepsilon_o E)^{1/2}. \quad (2.5)$$

Using this expression in (2.4) yields

$$N = 4\pi V (kT_c/hc)^3 K(0,1,u_c), \quad (2.6)$$

where

$$K(0,1,u_c) = \int_0^\infty \frac{(x^2 + 2u_c x)^{1/2}(x + u_c)}{e^x - 1} dx, \quad u_c = mc^2/kT_c.$$

Approximate solutions of (2.6) for each of the two limiting cases may be obtained by expanding the numerator of the integrand of $K(0,1,u_c)$ to give

$$K(0,1,u_c) \approx (\pi u_c/2)^{1/2} \zeta(3/2) u_c, \text{ if } u_c \gg 1 \text{ (nonrelativistic)}$$

and

$$K(0,1,u_c) \approx 2\zeta(3), \text{ if } u_c \ll 1 \text{ (extreme relativistic)}.$$

In these two limiting cases, the relevant condensation temperature is given by

$$kT_c = \frac{h^2}{2\pi m} \left[\frac{N}{V\zeta(3/2)}\right]^{2/3}, \text{ (nonrelativistic)} \quad (2.7)$$

and

$$kT_c = \left[\frac{h^3 c^3 N}{8\pi V \zeta(3)}\right]^{1/3}, \text{ (extreme relativistic)}. \quad (2.8)$$

As is shown in detail in reference [2], if condensation temperature is plotted against mass for each of the above limiting cases, the two resulting straight lines intersect on another straight line whose equation is given by eliminating $N/V$ between (2.7) and (2.8). The result of this elimination is the equation

$$T_c = \frac{mc^2}{2k}\left\{\pi\left[\frac{\zeta(3/2)}{\zeta(3)}\right]\right\}^{1/3} = 8\times 10^{36} m \text{ (°K)}$$

Above this line relativistic effects become important. It might be noted also that the higher the concentration, the higher the rest mass at which these effects begin to appear. For the kaons mentioned above, this equation gives a value for the condensation temperature of $7.13 \times 10^{12}$ °K. Using this value in either (2.7) or (2.8) shows that, for these kaons, relativistic effects will come into play for values of the concentration above approximately $3 \times 10^{39}$, which implies above a mass density of approximately $2.67 \times 10^{15}$ gms/cc.

All of the above, apart from the brief application to kaons, is well documented in [2], from which further details may be extracted. However, to reach the above results, even in a more rigorous deduction, the condensation temperature is said to have been reached when $\alpha = 0$. It is claimed frequently, and correctly, that $\alpha$ must remain non-zero since, if not, the implication is that the number of particles in the lowest energy level becomes infinite. Initially, though, the total number of particles in the system is always assumed both fixed and finite. In most traditional approaches to the topic, as here, these latter points are either ignored or glossed over.

## 3. An alternative approach to Bose-Einstein condensation.

The so-called Bose-Einstein distribution, which is really an expression for the average number of particles in a given energy range as a function of the absolute temperature rather than a genuine distribution, may be derived from the negative binomial distribution

$$f_r^G(n; p) = \binom{\Omega + n - 1}{n} p^\Omega q^n \quad (3.1)$$

where $p$ and $q$ are the a priori occupation and absence probabilities, by writing it as a law of error for which the most probable value coincides with the mean of the distribution. The entropy is the potential function which determines the law of error and the second law of thermodynamics then determines the Bose-Einstein distribution. The negative binomial distribution (3.1) is actually the probability of finding $n$ identical unnumbered particles in $\Omega$ boxes, with no empty boxes. The binomial coefficient is the number of ways the $n$ particles may be distributed among the $\Omega$ boxes. If $\Omega$ equals unity. (3.1) becomes

$$f^L(n;p) = pq^n, \quad (3.2)$$

which is the geometric distribution. Hence, the case where the density of states is unity does **not** belong to the negative binomial distribution but, rather, to the geometric distribution. This seemingly innocuous result immediately raises questions concerning the extraction of the term relating to the number of particles in the lowest energy level from the sum and subsequently writing the expression for the total number of particles in the system as in (2.1).

However, as was shown some time ago [3], probability distributions belonging to the same family may enter into 'equilibrium' with one another in much the same way as material phases do, but can this idea be used in the present context? It does seem that the Bose-Einstein gas might be thought to owe its condensation phenomenon to an osmotic equilibrium established between the negative binomial and geometric distributions. The osmotic equilibrium that is maintained by a pressure difference between the two phases may be thought due to some semi-permeable 'membrane'. This 'membrane' would be permeable to the Bose-Einstein gas (particles in excited states) but **not** to the condensate (particles in the zero energy state). The Bose-Einstein condensation, or osmotic equilibrium, would exist at any temperature. However, at high temperatures, few particles will be in the zero energy state and so, the equilibrium will be insignificant. The onset of what is normally called Bose-Einstein condensation will occur at the point where the osmotic pressure exhibits a minimum with respect to the mole fraction.

The entropies

$$S^G\left(\overline{n}^G\right) = \left(\Omega + \overline{n}^G\right)\ln\left(\Omega + \overline{n}^G\right) - \overline{n}^G \ln \overline{n}^G - \Omega \ln \Omega \quad (3.3)$$

and

$$S^L(\bar{n}^L) = (1 + \bar{n}^L)\ln(1 + \bar{n}^L) - \bar{n}^L \ln \bar{n}^L \quad (3.4)$$

are found by casting the probability distributions (3.1) and (3.2) respectively as laws of error for which the most likely value of $n$ equals the mean of the distribution. If their derivatives at constant volume with respect to $\bar{n}^G$ and $\bar{n}^L$ respectively are equated to the corresponding expressions obtained from the second law of thermodynamics, then

$$\bar{n}^G(T) = \frac{\Omega}{z_G^{-1} e^{\varepsilon/kT} - 1} \quad (3.5)$$

and

$$\bar{n}^L(T) = \frac{z_L}{1 - z_L}, \quad (3.6)$$

where the fugacities are defined by $z_G = e^{\mu_G/kT} = e^{\alpha_G}$ and $z_L = e^{\mu_L/kT} = e^{\alpha_L}$, result. For (3.6) to be positive and finite, the chemical potential $\mu_L$ must be finite and less than zero. The Bose-Einstein condensation is due, then, to the osmotic equilibrium established between the two phases.

The latter two equations give the mean numbers for the two phases with chemical potentials $\mu_G$ and $\mu_L$. For chemical equilibrium these must be equal and, by varying the common temperature and pressures $P^G$ and $P^L$ of the two phases, the maintenance of equilibrium implies

$$V_m^G dP^G - V_m^L dP^L = (S_m^G - S_m^L) dT, \quad (3.7)$$

where $T(S_m^G - S_m^L)$ is the latent heat of transition. $S_m^G$ and $S_m^L$ are the molar entropies and $V_m^G = V/\bar{n}^G$ and $V_m^L = V/\bar{n}^L$ are the specific molar volumes of the two phases. Using the Gibbs' relation, the pressures of the two phases are found to be

$$P^G(T) = -\frac{\Omega T}{V} \ln(1 - z_G e^{-\varepsilon/T}) \quad (3.8)$$

and

$$P^L(T) = -\frac{T}{V} \ln(1 - z_L). \quad (3.9)$$

These pressures may be seen to satisfy the generalised Clapeyron equation (3.7), establishing the fact that a phase equilibrium has been attained. This equilibrium is independent of the actual nature of the system, which is specified only when a definite expression is given for $\Omega$, the number of states in a given energy interval. Hence, the phase equilibrium may apply to relativistic particles of zero mass as well as to non-relativistic ones. The stability criteria must be obtained in terms of the total pressures but, since they must be independent of the energy of a particular mode, the phase equilibrium must be established between each excited mode belonging to the negative binomial distribution and the ground state belonging to the geometric distribution. This must also be so since the excited modes do not interact with one another.

At the point where the above pressures are equal, $\mu_G > \mu_L$. Hence, in order to achieve equilibrium, $P^L > P^G$. The excess pressure, $P^L - P^G$, is the osmotic pressure. The total osmotic pressure is

$$\Pi_{tot} = -\frac{T}{V}\ln(1-z) - \frac{T}{\lambda^3} g_{5/2}(z), \quad (3.10)$$

where $g_n(z) = \frac{1}{\Gamma(n)} \int \frac{x^{n-1} dx}{z^{-1} e^x - 1}$, $\lambda^3 = h^3 (2\pi mkT)^{-3/2}$ and the second term in (3.10) is obtained by integrating (3.8) over all energy states. The establishment of an osmotic equilibrium with a positive osmotic pressure requires $P^L = -T \ln(1-z)/V$, the pressure of the condensate, to be greater than $P_{tot}^G = T g_{5/2}(z)/V$, the total gas pressure. This will occur for values of $z$ very close to unity.

The chemical potential is always related to both the temperature and volume via the relation

$$N = \frac{z}{1-z} + \frac{V}{\lambda^3} g_{3/2}(z), \quad (3.11)$$

since the total number of particles is conserved. Hence, it cannot be concluded, in a truly realistic model, that the total gas pressure is independent of the volume below the critical temperature; - this conclusion follows only if $\mu = 0$ and, as is seen from (3.11), this would imply

$$\frac{z}{1-z} \to \infty$$

and would make a nonsense of the idea of particle number conservation.

The critical point is determined by requiring the total osmotic pressure to be a minimum with respect to the mole fraction. By writing

$$N = \bar{n}^L + \bar{n}_{tot}^G$$

and using

$$g_{n-1} = z \frac{\partial g_n(z)}{\partial z} \quad \text{and} \quad z = \frac{\bar{n}^L}{1 + \bar{n}^L},$$

(3.10) may be written

$$\frac{V}{T} \Pi_{tot} = (N+2) \ln\left(1 + \bar{n}^L\right) - N \ln \bar{n}^L. \quad (3.12)$$

The critical temperature is determined by

$$\left(\partial \Pi_{tot} / \partial \bar{n}^L\right)_T = 0.$$

This leads to the condition

$$\bar{n}^L = N/2 \quad \text{or} \quad \bar{n}_{tot}^G = \bar{n}^L,$$

where

$$\bar{n}^L = \frac{z}{1-z} \quad \text{and} \quad \bar{n}_{tot}^G = \frac{V}{\lambda^3} g_{3/2}(z).$$

It follows that, at the critical point

$$z_c = \frac{N}{N+2},$$

which is almost unity for $N \gg 1$ and $\Pi_{tot}$ is a minimum.
Hence, at the critical point

$$\frac{N}{2} = \bar{n}^L = \bar{n}_{tot}^G = \frac{V}{\lambda_c^3} g_{3/2}(z_c) \approx \frac{V}{\lambda_c^3} g_{3/2}(1).$$

This leads to a value for the critical temperature of a non-relativistic Bose-Einstein gas of 1.98 °K. However, the true critical temperature will be slightly higher than this since $g_{3/2}(z)$ is a bounded, positive, monotonically increasing function of $z$ between zero and one.

This approach is obviously quite appealing, producing, as it does, a figure for the condensation temperature so close to the experimentally measured value of 2.18 °K as opposed to that derived by traditional means. Also, and very importantly, the traditional approach ends up with an infinite number of particles in the ground state and this hardly compatible with the assumption of a fixed number of particles. As far as applications to possible Bose condensates in stars are concerned, the new approach has little effect on the final value for the condensation temperature. In fact, the difference between the two approaches is seen to amount to introducing a factor $(1/2)^{2/3}$ into the final expression for the critical temperature and this would mean that the condensation temperature for the kaons mentioned above would be $4.49 \times 10^{12}$ °K, rather than $7.13 \times 10^{12}$ °K as given earlier.

## 4. Questions remaining.

Even with this alternative approach to Bose-Einstein condensation, problems remain. As stated earlier, the total osmotic pressure $\Pi_{tot}$, as given by (3.10), will be positive for values of $z$ very close to unity. However, the idea presented here is that the critical temperature occurs when

$$z = z_c = \frac{N}{N=2}$$

and, at this value of $z$, if $N = 10^{24}$, $-\ln(1-z) \approx 55.26$, which is not large and is appreciably smaller than $Vg_{5/2}(z)/\lambda^3$.

Two possibilities immediately occur as resolutions to this difficulty:
(a) since the stability criteria must be independent of the energy of a particular mode, as mentioned earlier, the phase equilibrium must be established between each excited mode belonging to the negative binomial distribution and the ground state belonging to the geometric distribution. It is possible for $P^L > P^G$ for all excited states without $P^L$ being greater than $P^G_{tot}$; that is, without $\Pi_{tot}$ being positive;
(b) if the number of states at $\varepsilon = 0$ is not one but $\omega$, where $\omega$ may be large,

$$z = z_c = \frac{N}{N + 2\omega}$$

and

$$1 - z_c = \frac{2\omega}{N + 2\omega}.$$

If $\omega$ is large enough, $-\ln(1-z_c)$ can become greater than $Vg_{5/2}(z_c)/\lambda_c^3$. However, if this is the correct route towards the resolution of the problem, (3.2) would be replaced by an expression derived by putting $\Omega = \omega$ in (3.1) not equal to one, and this would give

$$f^L(n:p) = \binom{\omega + n - 1}{n} p^\omega q^n,$$

that is, a second negative binomial distribution. The problem would then be to examine an osmotic equilibrium established between two negative binomial distributions.

These latter points obviously require more time and thought devoted to them. However, more details may be found in reference [3] cited below, where further references are given also.

**References.**